\documentclass[authoryear]{elsarticle}
\makeatletter
\def\ps@pprintTitle{%
 \let\@oddhead\@empty
 \let\@evenhead\@empty
 \def\@oddfoot{}%
 \let\@evenfoot\@oddfoot}
\makeatother
\usepackage[margin=2.5cm]{geometry}
\usepackage{caption}
\usepackage[utf8]{inputenc}
\usepackage{graphicx}
\usepackage{physics}
\usepackage{longtable}
\usepackage{tabularx}
\usepackage{xltabular}
\usepackage{csquotes}
\usepackage[title]{appendix}
\graphicspath{ {./images/} }
\usepackage{setspace}
\usepackage{amsmath}
\usepackage{multirow}
\BeforeBeginEnvironment{align}{\begin{singlespacing}}
\AfterEndEnvironment{align}{\end{singlespacing}}

\BeforeBeginEnvironment{align*}{\begin{singlespacing}}
\AfterEndEnvironment{align*}{\end{singlespacing}}

\newcolumntype{L}[1]{>{\centering\arraybackslash\hsize=#1\hsize}X}

\newcolumntype{M}[1]{>{\arraybackslash\hsize=#1\hsize}X}

\usepackage{verbatim}

\newcommand{\detailtexcount}[1]{%
  \immediate\write18{texcount -merge -sum -q #1.tex output.bbl > #1.wcdetail }%
  \verbatiminput{#1.wcdetail}%
}

\begin{document}
\begin{frontmatter} 


\title{Stock Price Predictability and the Business Cycle
via Machine Learning}

\author[1]{Li Rong Wang}
\ead{LWANG039@e.ntu.edu.sg}

\author[2]{Hsuan Fu\corref{cor1}}
\ead{hsuan.fu@fsa.ulaval.ca}

\author[1]{Xiuyi Fan}
\ead{xyfan@ntu.edu.sg}

\cortext[cor1]{Corresponding author}

\address[1]{School of Computer Science and Engineering, Nanyang Technological University,
Singapore}

\address[2]{Department of Finance, Insurance and
Real Estate, Université Laval,
Canada}

\begin{abstract} 

We study the impacts of business cycles on machine learning (ML) predictions. Using the S\&P 500 index, we find that ML models perform worse during most recessions, and the inclusion of recession history or the risk-free rate does not necessarily improve their performance. Investigating recessions where models perform well, we find that they exhibit lower market volatility than other recessions. This implies that the improved performance is not due to the merit of ML methods but rather factors such as effective monetary policies that stabilized the market. We recommend that ML practitioners evaluate their models during both recessions and expansions.

\end{abstract}

\begin{keyword}
Empirical Asset Pricing \sep Business Cycle \sep Machine Learning \sep Long Short-Term Memory \sep Gated Recurrent Units \sep S\&P 500

\JEL C52 \sep C58 \sep G12 \sep G17
\end{keyword}

\end{frontmatter}  

\newpage
\section{Introduction} 


The relationship between the business cycle and stock market volatility has been extensively studied in the literature. It is an issue of great importance for policy and investment decision makers \citep{schwert1989business,fama1990stock,corradi2013macroeconomic,chauvet2013realized}. Empirical studies have been used to examine whether stock market volatility, which behaves differently in expansion and recession periods, can be predicted by macroeconomic variables \citep{schwert1989business,hamilton1996stock}. Research has also established a link between stock market volatility and macroeconomic fundamentals \citep{engle2008spline,diebold2008macroeconomic,corradi2013macroeconomic,choudhry2016stock}. However, despite recent successes in developing machine learning (ML) models for predicting financial prices of different assets (see e.g., \cite{guetal2020, heaton2017portfolioselection, gu2021autoencoder, bianchi2021bond}), there is little literature discussing the impact of business cycles and market volatility on stock price forecasting with ML models. This paper fills this gap and explores the data-shifting effects of market volatility resulted from recessions on ML models. Specifically, we focus on answering the following three research questions in this work: 

\begin{enumerate}
    \item Do ML models perform differently during the recession compared to non-recession?
    \item Does including recession data in the in-sample (training) period improve ML performance?
    \item Does the risk-free rate predictor improve ML performance during recessions?
\end{enumerate}

We situate our study in forecasting S\&P 500 price index with a spectrum of ML models that are commonly used for time series prediction such as the Long short-term memory (LSTM), bidirectional LSTM (BLSTM), and Gated Recurrent Unit (GRU). These models have different Recurrent Neural Network (RNN) layers and are used in financial forecasting in the recent literature \citep{seabe2023forecasting,hamayel2021novel,cao2020deep}. We study model forecasting performance in all past seven US recessions from 1969 to 2020 and build multiple forecasting models for each period. We conjecture that predictions are less accurate during recession periods due to the increased market volatility than during expansion periods. We aim to offer some insights on the best possible data set for training an ML model. 


Our main finding mostly confirms the conjecture. For the majority of models, we report larger forecasting errors in recessions than in expansions. 
However, a few models do not coincide with our conjecture. We observe that these models were all evaluated on the two recessions in the late 70s. Note that the oil crisis with high inflation in the late 70s resulted in a major change in monetary policy-making. We argue that this explains the low volatility in the stock market during these periods. Thus, these two recessions should be considered exceptions in our analysis. 
Secondly, we have added the recession observations into the in-sample (training) set and find that half of the models (22 out of 42) show signs of improved forecasting performance. However, there is no clear pattern on whether a specific type of models (e.g., LSTM, GRU or BLSTM) or some specific periods that benefit from such inclusion. Hence, the benefit of including recession data remains opaque for forecasting in recessions. 
Thirdly, we have also experimented with introducing an additional input, the risk-free rate approximated by the 3-months US treasury bill interest rate, as another attempt to improve model performance. In our experiment, 25 out of 42 models show an increase in forecasting accuracy after the risk-free rate were introduced. We interpret this as the inclusion of risk-free rate may be beneficial for stock price forecasting, but the effect is not consistent. To ensure the robustness of our findings, in addition to standard prediction errors (mean squared error), we have also used R-Squared and Certainty Equivalent Return Gain (CERG) as our evaluation metrics. We have found results from all three metrics are well aligned. 

This paper is related to both understanding the relationship between the business cycles and market volatility as well as the fast-growing ML literature in finance. The contribution is two-fold. First, we explore the macroeconomic aspect of the stock price dynamics by examining the predictability during recession versus expansion periods. Second, we demonstrate that poor prediction performance during a recession cannot simply be mitigated by enriching the information content of the in-sample set. 

The rest of the paper is organized as follows. Section \ref{sec:m} reviews the methodology, detailing the data, model and evaluation metrics used. Section \ref{sec:r} presents and interprets the prediction results from the ML models. Section \ref{sec:d} discusses the economic implications 
and offers some concluding remarks.

\section{Methodology}\label{sec:m} 


\subsection{Data Set} \label{sec:data_split} 
\begin{figure}[h]
\centering
\includegraphics[trim=10 25 10 10,clip,width=\textwidth]{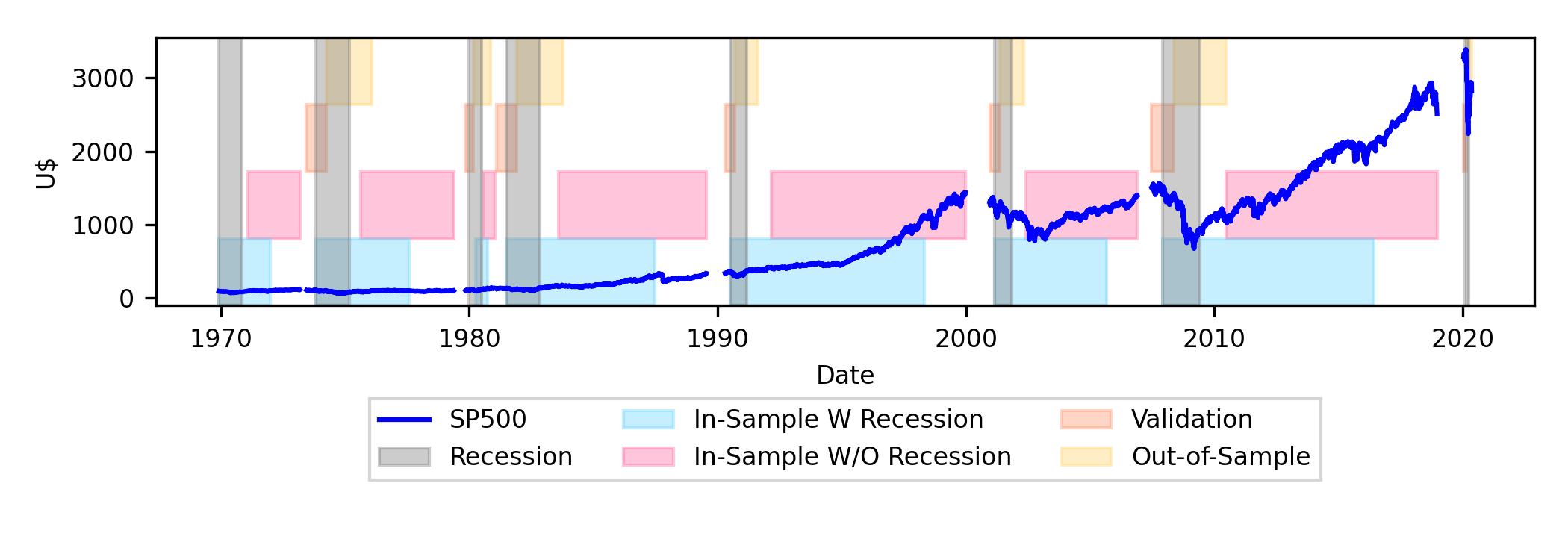}
\caption[In-Sample, Validation and Out-of-Sample Data Splits. The figure shows the S\&P 500 composite price index from 1 December 1969 to 13 May 2020.]{In-Sample (training), Validation and Out-of-Sample (testing) Data Splits. The figure shows the S\&P 500 price index from 1 December 1969 to 13 May 2020 (blue line). Greyed-out regions show periods of National Bureau of Economic Research (NBER) recession. Blue, pink, orange, and yellow regions correspond to ``In-Sample With Recession (ISWR)", ``In-Sample Without Recession (ISWOR)", ``Validation" and ``Out-of-Sample" data parts. Each sub-period consists of 4 consecutive data parts. ``ISWR" and ``ISWOR" refer to in-sample sets with both non-recession and recession data and with only non-recession data respectively. ``Validation" is used for hyper-parameter tuning. ``Out-of-Sample" (OOS) is used for model evaluation.}
\label{fig:data_split}
\vspace{-10pt}
\end{figure}

The forecasting target is S\&P 500 price index. The inputs are S\&P 500 index (up to one day before the forecasting date), the Fama-French 3 Factors, Momentum Factor \citep{fama1993common, fama2017choosing}, and the Risk-free rate (for answering research question 3). The data spans from 1 December 1969 to 13 May 2020 and it is divided into 7 overlapping sub-periods (Aligned with the number of recessions by NBER) as illustrated in Figure~\ref{fig:data_split}. Each sub-periods spans two adjacent recessions (resulting in sub-period with varying lengths). To maximize the use of the collected data, the validation and out-of-sample sets overlap with the in-sample set of the following sub-period. Each model was trained on ``In-Sample With Recession (ISWR)" or ``In-Sample Without Recession (ISWOR)", hyper-parameter-tuned with ``Validation" and evaluated on ``Out-of-Sample" from a single sub-period. In each sub-period, we have ensured that (1) ``Out-of-Sample" follows ``In-Sample" to prevent leakage, i.e. models trained with future data and evaluated on past data; (2) ISWR and ISWOR are of equal sizes to ensure models trained on either are comparable; and (3) ``Out-of-Sample" is equal parts recession and expansion data to  ensure evaluated recession and expansion performances are comparable, (4) 70\% and 30\% of the second recession in each sub-period are assorted into ``Out-of-Sample" and ``Validation" respectively, (5) a buffer region of 10\% of remaining data (between the first recession and ``Validation") was kept before and after ISWOR, (6) the proportion of recession data in ISWR is limited to 0.5.

\subsection{Model Architecture, Training Methods and Evaluation Metrics}\label{sec:model_architecture_n_training}  
\begin{figure}[t]
\centering
\includegraphics[width=\textwidth]{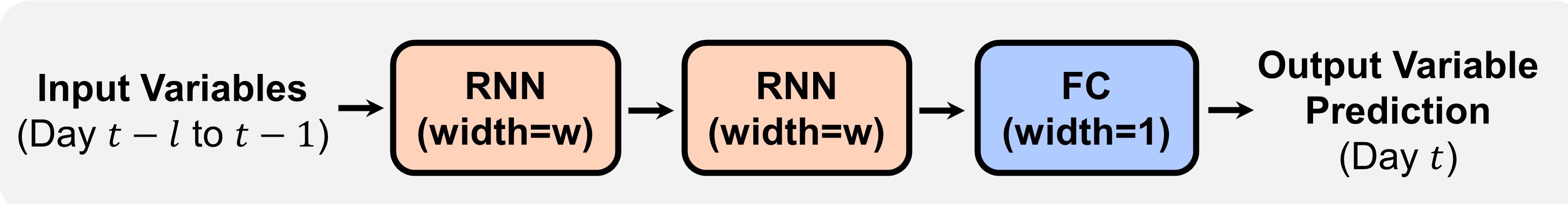}
\caption[Shared model architecture.]{Shared model architecture. Each model comprises 2 RNN layers of equal width \(w\) and 1 fully connected layer (FC) with width 1. The RNN layers are replaced with LSTM, bidirectional LSTM, and GRU layers to form 3 different models. To prevent over-fitting, we have used L2 regularisation (l2=0.01) and Dropout (rate=0.2) to RNN layers, as well as early stopping when validation error increases (Patience=10). For all models, we adopt cross-validation to select hyper-parameters for achieving the best performance for each model. In particular, width ($w \in [32, 64, 128]$) and time lag ($l \in [5, 7, 9]$) were tuned with grid-search. Evaluated models were trained with Adam optimizer (learning rate=0.001) using tuned hyper-parameters.}
\label{fig:model_architecture}
\vspace{-20pt}
\end{figure}

\doublespacing
Three sets of models sharing the same architecture \citep{cao2020deep} (Figure~\ref{fig:model_architecture}) but with different RNN layers are experimented. RNN layers (LSTM, BLSTM and GRU) learn temporal information from input sequences to predict future stock prices. LSTM layers store long-term and short-term information in their cell and hidden states. This information is controlled by forget ($f(t)$), input ($i(t)$) and output ($o(t)$) gates \citep{hochreiter1997long}. A relu \citep{agarap2018deep} LSTM cell can be described as follows:

\vspace{-15pt}
\begin{align}
f(t)&=\sigma(U_f^T x(t)+W_f^T h(t-1)+b_f)   \\ 
i(t)&=\sigma(U_i^T x(t)+W_i^T h(t-1)+b_i) \\
o(t)&=\sigma(U_o^T x(t)+W_o^T h(t-1)+b_0 ) \\
\tilde c(t)&=relu(Uc^T x(t)+W_c^T h(t-1)+b_c ) \\
c(t)&=\tilde c(t)*i(t)+c(t-1)*f(t)) \\
h(t)&=relu(\tilde c(t))* o(t)
\end{align}
\noindent
where,
\vspace{-20pt}
\begin{align*}
f(t) &= \text{Forget gate at time t,} &
i(t) &= \text{Input gate at time t,} &
o(t) &= \text{Output gate at time t,}\\
x(t) &= \text{Features at time t,} &
h(t) &= \text{Hidden state at time t,} &
c(t) &= \text{Cell state at time t,}\\
\tilde c(t) &= \text{Candidate values at time t,}&
U^T_j, W^T_j &= \text{Weight matrix for j,} &
b_j &= \text{Bias for j,} \\
\sigma() &= \text{Sigmoid function,} &
relu() &= \text{Relu function,} &
* &= \text{Elementwise Multiplication.}
\end{align*}

\doublespacing
The two LSTM layers in BLSTM read inputs in the forward direction (like LSTM) and in the reverse direction. This preserves past and future information \citep{1556215}. Unlike LSTM, GRU layers use 2 gates (reset and update) and store both long-term and short-term information in its hidden state \citep{cho2014properties}. The following equations describe computations in a relu GRU cell:

\vspace{-20pt}
\begin{align}
r(t)&=\sigma(U_r^T x(t)+W_r^T h(t-1)+b_r)\\
z(t)&=\sigma(U_z^T x(t)+W_z^T h(t-1)+b_z) \\
h'(t)&=relu(U_h^T x(t)+W_h^T(r(t)*h(t-1))+b_h)\\
h(t)&=z(t)*h(t-1)+(1-z(t))*h'(t)
\end{align}
\noindent
where,
\begin{singlespace}
\vspace{-20pt}
\begin{equation*}
\begin{aligned}[c]
r(t) &= \text{Reset gate at time t,}&
z(t) &= \text{Update gate at time t,}&
h'(t) &= \text{Candidate activation at time t.}\\
\end{aligned}
\end{equation*}
\end{singlespace}
\vspace{-5pt}
\noindent
While RNN layers read inputs sequentially, FC layers read inputs, \(x\), concurrently. The output, \(y\), is computed with: \(\vectorbold{\hat y} = \vectorbold{W}^T \vectorbold{x} + \vectorbold{b}\) where \(\vectorbold{W}\) is the weight matrix and \(\vectorbold{b}\) is the bias vector. 
%

All models are evaluated with three metrics: Mean Squared Error (MSE), R-Squared \(R^2\), and Certainty Equivalent Return Gain (CERG) as shown in Table~\ref{tab:metric_summary}. An improvement in OOS performance means a reduction in MSE or an increase in \(R^2\) and CERG.

\begin{small}
\begin{singlespace}
\begin{xltabular}{\textwidth}{|M{0.15}|M{0.3}|M{0.55}|}    
\caption{Summary of evaluation metrics. MSE quantifies the deviation between true and predicted stock prices. Prior to MSE calculation, Z-score normalization (fitted on $Y_i$) is applied to $Y_i$ and $\hat Y_i$ to ensure cross-sub-period comparability. \(R^2\) measures the relative error of an Equity Risk Premium (ERP) forecasting model in relation to error from using historical average \citep{campbell2008oosrsquared}. CERG estimates the fee a mean-variance investor would pay to use the proposed model \citep{lu2022cer}. 
} \\ \hline
\multicolumn{1}{|c|}{\textbf{Metrics}} & 
\multicolumn{1}{c|}{\textbf{Formulas}} &
\multicolumn{1}{c|}{\textbf{Variable Definitions}}
\\ \hline 
\endfirsthead
\label{tab:metric_summary} 
Mean Squared Error (MSE) & 
$MSE = 1/n \sum^n_{i=1} (Y_i-\hat Y_i)^2$ &
    \begin{tabular}{rl}
    $Y_i$:& True stock price \\
    $\hat Y_i$:& Predicted stock price \\
    $n$:& Size of OOS set \\
    \end{tabular} 
    \\ \hline
R-Squared $R^2$ & 
$R^2 = 1- \frac{\sum\epsilon_t^2}{\sum v_t^2}$ &
    \begin{tabular}{rl}
    $\epsilon_t$:& \(ERP_{true} - ERP_{predicted}\) at $t$. \\
    $v_t$:& \(ERP_{true} - ERP_{historical\_average}\) at $t$. \\
    $ERP_{predicted}$:& $log(\frac{\hat S_t}{S_{t-1}}) - rf_{t-1}$ \\
    $ERP_{true}$:& $log(\frac{S_t}{S_{t-1}}) - rf_t$ \\
    $S_t$:& Stock price at $t$. \\
    $rf_t$:& Risk-free rate at $t$. \\
    \end{tabular} 
    \\ \hline
Certainty Equivalent Return Gain (CERG) & 
$\Delta = \hat U_p - \hat U_b$ &
    \begin{tabular}{rl}
    $\hat U_p$:& \( mean(\hat R_t)-\frac{\gamma}{2}Var(\hat R_t)\) over OOS period.\\
    $\hat U_b$:& \( mean(R_t)-\frac{\gamma}{2}Var(R_t)\) over data before OOS.\\
    $\hat R_t$:& \(log(\frac{\hat S_t}{S_{t-1}})\).\\
    $R_t$:& \(log(\frac{S_t}{S_{t-1}})\).\\
    $S_t$:& Stock price at time $t$. \\
    $\gamma$:& Relative risk aversion coefficient = 3. \\
    \end{tabular} 
    \\ \hline
\end{xltabular}
\end{singlespace}    
\end{small}

\section{Experiment Results}\label{sec:r} 
%
\subsection{OOS Performance During Recession and Non-Recession}
\label{sec:recession_non_recession_perf} 


\begin{figure}[htbp]
    \centering
    \includegraphics[width=\textwidth]{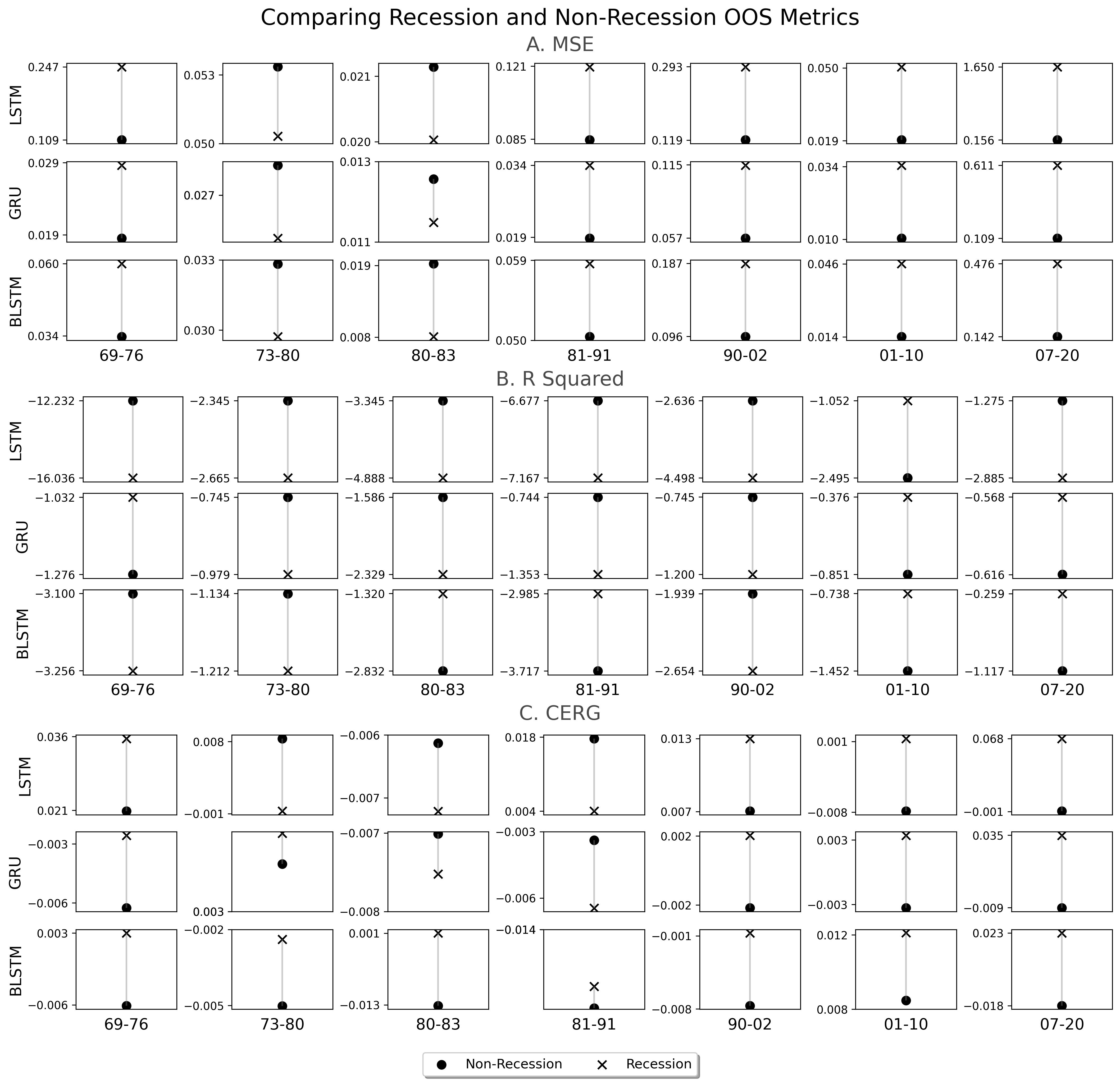}
    \caption[Lollipop chart comparing the Mean Squared Errors (MSEs), R-Squared values and CER Gains of stock price forecasting models during Recession and Non-Recession OOS.] {Lollipop chart comparing the Mean Squared Errors (MSEs), R-Squared values and CER Gains of stock price forecasting models during Recession and Non-Recession OOS. The figure displays the MSEs, \(R^2\) and CERG of each of the 21 ML models on the recession (Marked with crosses) and non-recession OOS set (Marked with circles). The figure shows a greater proportion of models (15 out of 21 models) having recession MSE $>$ non-recession MSE. Comparing R-Squared values, a greater proportion of models (13 out of 21 models) have recession R-Squared $<$ non-recession R-Squared. The figure also shows a greater proportion of models (16 out of 21 models) having recession CER Gains $>$ non-recession CER Gains. This strongly suggests that ML models are susceptible to data shifts caused by the business cycle.}
    \label{fig:re_non_re_oos}
\end{figure}

We have trained 21 models on non-recession data and compared their performance on both recession and non-recession out-of-sample data. Their results support the hypothesis that recessions have degenerative effects on forecasting due to data shifts resulted from the increased market volatility during recessions. As shown in Figure~\ref{fig:re_non_re_oos}.A, most of the models had higher mean squared error during a recession, as expected. However, six models did not align with this hypothesis, and further investigation revealed that these models were concentrated in two sub-periods coinciding with recessions caused by the oil crisis and high inflation. This period also saw the US Federal Reserve implement rule-based monetary policy, which may have led to a structural change in the economy with opposite implications from other recession periods, resulting in the observed discrepancy. 

\begin{figure}[htbp]
\centering
\includegraphics[width=\textwidth]{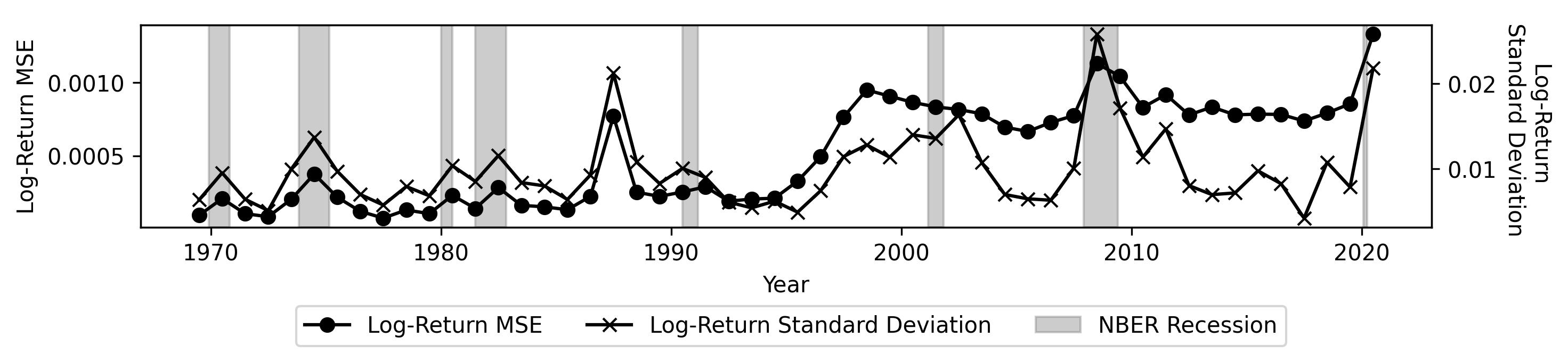}
\caption[Graph depicting yearly log-return MSE (Mean Squared Error) of GRU models and log-return standard deviations.]{Graph depicting yearly log-return MSE (Mean Squared Error) of GRU models and log-return standard deviations. Log-return MSE refers to the mean squared deviation between predicted and actual log-return values. We observed higher volatility and MSE during a recession and a high positive correlation between MSE and volatility. 
}
\label{fig:yearly_se_sd}
\vspace{-10pt}
\end{figure}

Studying the OOS volatility of all sub-periods (Figure \ref{fig:yearly_se_sd}), we observe that recession periods are intrinsically more volatile than non-recession periods. Focusing on the two sub-periods (73-80 and 80-83), we observe similar volatility between the recession and non-recession OOS data. This potentially implies that the effective monetary policy could have stabilized the stock market during both recessions. Hence, resulting in similar predictive performance during recession and non-recession. Figure \ref{fig:yearly_se_sd} also demonstrates a strong correlation between stock market volatility and prediction performance (Peaks in SE coincide with peaks in volatility). This suggests that the volatile nature of most recessions could have caused the observed performance deterioration.

Comparing the \(R^2\) during recession and non-recession, a similar trend was observed; a greater proportion of models (13) had lowered \(R^2\) during a recession (Figure \ref{fig:re_non_re_oos}.B), also supporting our hypothesis. The two outlier recessions pointed out earlier was not observed with \(R^2\). This showcases \(R^2\)'s ability to normalise model performance according to prediction difficulty. Analyzing the CERG of 21 models during recession and non-recession (Figure \ref{fig:re_non_re_oos}.C), we observed that most models showed higher CERG during a recession (16). This observation is consistent with other studies; During periods of recession, expected returns are likely to be higher due to heightened levels of risk aversion \citep{fama1989business, lu2022cer}. 

\subsection{Effects of Including Recession Data in the In-sample Period} 
Having established that recessions worsen prediction accuracy, we will attempt to improve recession predictions  by changing the in-sample set information content. To this end, we enrich the in-sample set with recession data points. We have trained 42 models (Recession-trained (RT) or only Non-Recession-Trained (NRT); 7 sub-periods; 3 model types). Comparing recession performances of RT and NRT models with the same sub-period and model type (21 comparisons) (Tables \ref{tab:re_non_re_trained_mse} and \ref{tab:re_non_re_trained_r2_cer}), we did not observe the majority of RT models improving from their NRT counterparts across all metrics (MSE, \(R^2\), CERG). This suggests that the inclusion of recession data in the in-sample set does not necessarily improve recession performance. This could be attributed to the diverse causes of a recession; different recessions have dissimilar stock market trends and hence one recession's stock prices are unhelpful when predicting a different recession's stock prices. For instance, in the 73-80 sub-period none of the RT models showed improvements in MSE or \(R^2\). In this sub-period, RT models were trained with the 1973 oil embargo recession and evaluated on the 1980 recession which was caused by the strict monetary policy in the US. Both recessions had vastly different causes and economic trends which likely resulted in the lacklustre improvements in recession prediction performance. 

\begin{small}
\begin{singlespace}
\begin{xltabular}{\textwidth}{*{6}{>{\centering\arraybackslash}X}}
    \caption[Table comparing the OOS Recession and Non-Recession MSEs of Non-Recession-Trained (NRT)  and Recession-Trained (RT) ML Models. ]{Table comparing the OOS Recession and Non-Recession MSEs of Non-Recession-Trained (NRT) and Recession-Trained (RT) ML Models. Half of all models (11 out of 21 models) have lower recession MSEs than NRT models. Comparing the non-recession MSEs of NRT and RT models, we observed improvements in MSEs with recession training data (16 out of 21 models). This table shows that the inclusion of recession data during training improves non-recession performance but not necessarily recession performance.} \\ \hline 
    \label{tab:re_non_re_trained_mse}
    \textbf{No.} & \textbf{Sub-period} & \textbf{Model Type} & \textbf{NRT MSE (1)} & \textbf{RT MSE (2)} & \textbf{(1) $>$ (2)} \\ \hline 
    \endfirsthead
    \endhead
    1 & 69-76 & BLSTM & 0.06 & 0.0525 & True \\ 
        2 & 69-76 & GRU & 0.0286 & 0.0282 & True \\ 
        3 & 69-76 & LSTM & 0.2469 & 0.0855 & True \\ 
        4 & 73-80 & BLSTM & 0.0297 & 0.0416 & False \\ 
        5 & 73-80 & GRU & 0.0268 & 0.0276 & False \\ 
        6 & 73-80 & LSTM & 0.0503 & 0.0662 & False \\ 
        7 & 80-83 & BLSTM & 0.0081 & 0.0126 & False \\ 
        8 & 80-83 & GRU & 0.0115 & 0.011 & True \\ 
        9 & 80-83 & LSTM & 0.02 & 0.0292 & False \\ 
        10 & 81-91 & BLSTM & 0.0586 & 0.0349 & True \\ 
        11 & 81-91 & GRU & 0.034 & 0.038 & False \\ 
        12 & 81-91 & LSTM & 0.1207 & 0.0869 & True \\
        13 & 90-02 & BLSTM & 0.1867 & 0.1155 & True \\ 
        14 & 90-02 & GRU & 0.1147 & 0.1165 & False \\ 
        15 & 90-02 & LSTM & 0.2926 & 0.166 & True \\ 
        16 & 01-10 & BLSTM & 0.0461 & 0.053 & False \\ 
        17 & 01-10 & GRU & 0.0345 & 0.0344 & True \\ 
        18 & 01-10 & LSTM & 0.0505 & 0.0556 & False \\ 
        19 & 07-20 & BLSTM & 0.4764 & 0.5318 & False \\ 
        20 & 07-20 & GRU & 0.6112 & 0.4516 & True \\ 
        21 & 07-20 & LSTM & 1.6498 & 1.2866 & True \\ \hline
    \multicolumn{3}{c}{Count} & 21.0 & 21.0 & 11 \\ \hline
\end{xltabular}
\vspace{-10pt}
\end{singlespace}
\begin{singlespace}
\begin{xltabular}{\textwidth}{L{0.4}L{0.9}L{1.0}L{1.1}L{1.1}L{0.9}L{1.1}L{1.1}L{0.9}}  
    \caption[Table comparing the OOS Recession and Non-Recession R-Squared ($R^2$) and CER Gains of Non-Recession-Trained (NRT)  and Recession-Trained (RT) ML Models.]{Table comparing the OOS Recession and Non-Recession R-Squared ($R^2$) and CER Gains of Non-Recession-Trained (NRT)  and Recession-Trained (RT) ML Models. Roughly half of all models (10/11 out of 21 models) have higher $R^2$ and CER gain than NRT models. Comparing the non-recession $R^2$s and CER gains of NRT and RT models, we observed improvements in $R^2$s and CER gains with recession training data (16 and 13 out of 21 models). This table shows that the inclusion of recession data during training improves non-recession performance but not necessarily recession performance.} \\ \hline
    \label{tab:re_non_re_trained_r2_cer}
    \textbf{No.} & \textbf{Sub-period} & \textbf{Model Type} & \textbf{NRT $R^2$ (1)} & \textbf{RT $R^2$ (2)} & \textbf{(2) $>$ (1)} & \textbf{NRT CER Gain (3)} & \textbf{RT CER Gain (4)} & \textbf{(4) $>$ (3)} \\ \hline
    \endfirsthead
    \endhead
        1 & 69-76 & BLSTM & -3.2564 & -2.8486 & True & 0.003 & -0.0053 & False \\ 
        2 & 69-76 & GRU & -1.0315 & -0.9863 & True & -0.0026 & 0.0003 & True \\ 
        3 & 69-76 & LSTM & -16.0357 & -5.324 & True & 0.0356 & -0.0096 & False \\ 
        4 & 73-80 & BLSTM & -1.2119 & -2.0695 & False & -0.0024 & 0.0074 & True \\ 
        5 & 73-80 & GRU & -0.9787 & -1.0372 & False & 0.0034 & 0.0047 & True \\ 
        6 & 73-80 & LSTM & -2.6653 & -4.0127 & False & -0.0007 & -0.0097 & False \\ 
        7 & 80-83 & BLSTM & -1.3205 & -2.6981 & False & 0.001 & -0.0071 & False \\ 
        8 & 80-83 & GRU & -2.3293 & -2.2333 & True & -0.0075 & 0.0082 & True \\ 
        9 & 80-83 & LSTM & -4.8877 & -7.8407 & False & -0.0072 & -0.0217 & False \\ 
        10 & 81-91 & BLSTM & -2.9847 & -1.4844 & True & -0.0143 & 0.0058 & True \\ 
        11 & 81-91 & GRU & -1.3532 & -1.6168 & False & -0.0065 & -0.0059 & True \\ 
        12 & 81-91 & LSTM & -7.1673 & -5.002 & True & 0.004 & -0.0097 & False \\ 
        13 & 90-02 & BLSTM & -2.6544 & -1.1502 & True & -0.0007 & 0.0036 & True \\ 
        14 & 90-02 & GRU & -1.2001 & -1.2376 & False & 0.002 & 0.0029 & True \\ 
        15 & 90-02 & LSTM & -4.4985 & -2.2247 & True & 0.013 & 0.0004 & False \\
        16 & 01-10 & BLSTM & -0.7382 & -1.1756 & False & 0.0121 & -0.0015 & False \\ 
        17 & 01-10 & GRU & -0.3763 & -0.3507 & True & 0.0034 & 0.0078 & True \\ 
        18 & 01-10 & LSTM & -1.0521 & -1.1505 & False & 0.0014 & 0.0094 & True \\ 
        19 & 07-20 & BLSTM & -0.2591 & -0.383 & False & 0.0229 & 0.0306 & True \\ 
        20 & 07-20 & GRU & -0.5681 & -0.1678 & True & 0.0349 & 0.0308 & False \\ 
        21 & 07-20 & LSTM & -2.885 & -3.3818 & False & 0.0679 & -0.0086 & False \\ \hline
        \multicolumn{3}{c}{Count} & 21.0 & 21.0 & 10 & 21.0 & 21.0 & 11 \\ \hline
\end{xltabular}
\end{singlespace}
\end{small}

\subsection{Effects of Adding Risk-Free Rate as an Input Variable} 

As we have identified that ML model performances are affected by market volatility, we explore whether using the risk-free rate ($r_f$) as a predictor would be beneficial for forecasting. We use the 3-months US treasury bill interest rate to proxy the risk-free rate. In Section \ref{sec:recession_non_recession_perf}, we observed 2 recession periods where all models performed better during the recession compared to non-recession. These 2 periods coincide with recessions caused by the implementation of monetary policies to manage elevated inflation. Since the risk-free rates are highly correlated with the policy rate announced by the Federal Reserve, we hypothesize that the inclusion of risk-free rates could improve recession performance, especially during a recession caused by strict monetary policies. To verify this, we have trained 84 models (Permutations of: With and without $r_f$, including or excluding recession data in the in-sample set, 7 sub-periods, 3 model types) and compared the recession performances of models trained with and without $r_f$. 

\begin{small}
\begin{singlespace}
\begin{xltabular}{\textwidth}{*{9}{>{\centering\arraybackslash}X}}
    \caption[Table comparing the OOS Recession MSEs of Non-Risk-Free-Trained (NRFT) and Risk-Free-Trained (RFT) ML Models. ]{Table comparing the OOS Recession MSEs of Non-Risk-Free-Trained (NRFT) and Risk-Free-Trained (RFT) ML Models. 9 out of 21 non-recession-trained models and 8 out of 21 recession-trained models showed a decrease in MSE after the risk-free rate was added as a feature. The above observation suggests that the risk-free rate may not be consistently useful for the prediction of stock prices.} \\ \hline
    \label{tab:rf_non_rf_trained_mse}
    \multirow{3}{*}{\textbf{No.}} & 
    \multirow{3}{*}{\textbf{Sub-period}} &
    \multirow{3}{1cm}{\textbf{Model Type}} &
    \multicolumn{3}{c}{\textbf{Non-Recession}} &
    \multicolumn{3}{c}{\textbf{Recession and Non-Recession}} \\
    \cline{4-6} \cline{7-9}
    & & & \textbf{NRFT MSE (1)} &
    \textbf{RFT MSE (2)} & \textbf{(1) $>$ (2)} & \textbf{NRFT MSE (1)} &
    \textbf{RFT MSE (2)} & \textbf{(1) $>$ (2)} \\ \hline
    \endfirsthead
    \endhead
    1 & 69-76 & BLSTM & 0.06 & 0.1101 & False & 0.0525 & 0.0528 & False \\ 
    2 & 69-76 & GRU & 0.0286 & 0.0301 & False & 0.0282 & 0.0281 & True \\ 
    3 & 69-76 & LSTM & 0.2469 & 0.2098 & True & 0.0855 & 0.0858 & False \\ 
    4 & 73-80 & BLSTM & 0.0297 & 0.0296 & True & 0.0416 & 0.0439 & False \\ 
    5 & 73-80 & GRU & 0.0268 & 0.0273 & False & 0.0276 & 0.0315 & False \\ 
    6 & 73-80 & LSTM & 0.0503 & 0.0504 & False & 0.0662 & 0.066 & True \\ 
    7 & 80-83 & BLSTM & 0.0081 & 0.0081 & False & 0.0126 & 0.0094 & True \\ 
    8 & 80-83 & GRU & 0.0115 & 0.0114 & True & 0.011 & 0.0113 & False \\ 
    9 & 80-83 & LSTM & 0.02 & 0.0203 & False & 0.0292 & 0.0193 & True \\ 
    10 & 81-91 & BLSTM & 0.0586 & 0.0597 & False & 0.0349 & 0.0522 & False \\ 
    11 & 81-91 & GRU & 0.034 & 0.0296 & True & 0.038 & 0.0378 & True \\ 
    12 & 81-91 & LSTM & 0.1207 & 0.1585 & False & 0.0869 & 0.0863 & True \\ 
    13 & 90-02 & BLSTM & 0.1867 & 0.1307 & True & 0.1155 & 0.1276 & False \\ 
    14 & 90-02 & GRU & 0.1147 & 0.1165 & False & 0.1165 & 0.1171 & False \\ 
    15 & 90-02 & LSTM & 0.2926 & 0.1541 & True & 0.166 & 0.2925 & False \\ 
    16 & 01-10 & BLSTM & 0.0461 & 0.0379 & True & 0.053 & 0.0389 & True \\ 
    17 & 01-10 & GRU & 0.0345 & 0.0352 & False & 0.0344 & 0.0341 & True \\ 
    18 & 01-10 & LSTM & 0.0505 & 0.1383 & False & 0.0556 & 0.1498 & False \\ 
    19 & 07-20 & BLSTM & 0.4764 & 1.5243 & False & 0.5318 & 0.8205 & False \\ 
    20 & 07-20 & GRU & 0.6112 & 0.5147 & True & 0.4516 & 0.7098 & False \\ 
    21 & 07-20 & LSTM & 1.6498 & 0.825 & True & 1.2866 & 2.182 & False \\ \hline
    \multicolumn{3}{c}{Count} & 21 & 21 & 9 & 21 & 21 & 8  \\ \hline
\end{xltabular}
\end{singlespace}
\begin{singlespace}
\vspace{-10pt}
\begin{xltabular}{\textwidth}{L{0.35}L{0.8}L{1.0}L{1.0}L{1.0}L{0.9}L{1.2}L{1.1}L{0.9}}   
    \caption[Table comparing the OOS Recession R-Squared ($R^2$), CER Gains of Non-Risk-Free-Trained (NRFT), and Risk-Free-Trained (RFT) ML Models.]{Table comparing the OOS Recession R-Squared ($R^2$), CER Gains of Non-Risk-Free-Trained (NRFT), and Risk-Free-Trained (RFT) ML Models. $R^2$ and CER gain gave similar conclusions to MSE for non-recession-trained models. 9 out of 21 models had an increase in $R^2$ after introducing risk-free rate and 11 out of 21 models had an increase in CER gain. The recession-trained models showed a different trend. A reduction in $R^2$ and an increase in CER gain were noted in most models after introducing risk-free rate.} \\ \hline
    \label{tab:rf_non_rf_trained_r2_cer}
    \textbf{No.} & \textbf{Sub-period} & \textbf{Model Type} & \textbf{NRFT $R^2$ (1)} & \textbf{RFT $R^2$ (2)} & \textbf{(2) $>$ (1)} & \textbf{NRFT CER Gain (3)} & \textbf{RFT CER Gain (4)} & \textbf{(4) $>$ (3)} \\ \hline 
    \endfirsthead
    \endhead
        \multicolumn{9}{c}{A. Trained With Only Non-Recession} \\ 
        1 & 69-76 & BLSTM & -3.2564 & -6.6875 & False & 0.003 & 0.03 & True \\ 
        2 & 69-76 & GRU & -1.0315 & -1.1063 & False & -0.0026 & 0.0025 & True \\ 
        3 & 69-76 & LSTM & -16.0357 & -13.302 & True & 0.0356 & 0.0405 & True \\ 
        4 & 73-80 & BLSTM & -1.2119 & -1.2049 & True & -0.0024 & -0.0023 & True \\ 
        5 & 73-80 & GRU & -0.9787 & -1.0147 & False & 0.0034 & 0.0039 & True \\ 
        6 & 73-80 & LSTM & -2.6653 & -2.6717 & False & -0.0007 & -0.0007 & False \\ 
        7 & 80-83 & BLSTM & -1.3205 & -1.3337 & False & 0.001 & 0.0016 & True \\
        8 & 80-83 & GRU & -2.3293 & -2.293 & True & -0.0075 & -0.0072 & True \\ 
        9 & 80-83 & LSTM & -4.8877 & -4.9694 & False & -0.0072 & -0.0081 & False \\ 
        10 & 81-91 & BLSTM & -2.9847 & -3.0602 & False & -0.0143 & -0.0146 & False \\ 
        11 & 81-91 & GRU & -1.3532 & -1.0737 & True & -0.0065 & -0.0017 & True \\ 
        12 & 81-91 & LSTM & -7.1673 & -10.0511 & False & 0.004 & 0.0191 & True \\ 
        13 & 90-02 & BLSTM & -2.6544 & -1.5077 & True & -0.0007 & -0.0043 & False \\ 
        14 & 90-02 & GRU & -1.2001 & -1.2378 & False & 0.002 & 0.002 & False \\ 
        15 & 90-02 & LSTM & -4.4985 & -1.8485 & True & 0.013 & 0.0098 & False \\ 
        16 & 01-10 & BLSTM & -0.7382 & -0.4996 & True & 0.0121 & 0.0045 & False \\ 
        17 & 01-10 & GRU & -0.3763 & -0.3914 & False & 0.0034 & 0.0048 & True \\
        18 & 01-10 & LSTM & -1.0521 & -4.3868 & False & 0.0014 & -0.0168 & False \\ 
        19 & 07-20 & BLSTM & -0.2591 & -2.7287 & False & 0.0229 & 0.0479 & True \\ 
        20 & 07-20 & GRU & -0.5681 & -0.3405 & True & 0.0349 & 0.0296 & False \\ 
        21 & 07-20 & LSTM & -2.885 & -1.1199 & True & 0.0679 & 0.031 & False \\ \hline
        \multicolumn{3}{c}{Count} & 21 & 21 & 9 & 21 & 21 & 11 \\ \hline
        \multicolumn{9}{c}{B. Trained With Non-Recession and Recession} \\ 
        1 & 69-76 & BLSTM & -2.8486 & -2.8941 & False & -0.0053 & -0.0061 & False \\ 
        2 & 69-76 & GRU & -0.9863 & -0.9865 & False & 0.0003 & -0.0002 & False \\
        3 & 69-76 & LSTM & -5.324 & -5.3563 & False & -0.0096 & -0.0102 & False \\ 
        4 & 73-80 & BLSTM & -2.0695 & -2.2303 & False & 0.0074 & 0.0083 & True \\ 
        5 & 73-80 & GRU & -1.0372 & -1.321 & False & 0.0047 & 0.0069 & True \\ 
        6 & 73-80 & LSTM & -4.0127 & -3.9926 & True & -0.0097 & -0.0095 & True \\ 
        7 & 80-83 & BLSTM & -2.6981 & -1.7467 & True & -0.0071 & 0.0002 & True \\ 
        8 & 80-83 & GRU & -2.2333 & -2.3283 & False & 0.0082 & 0.0088 & True \\ 
        9 & 80-83 & LSTM & -7.8407 & -4.7062 & True & -0.0217 & -0.0115 & True \\ 
        10 & 81-91 & BLSTM & -1.4844 & -2.5998 & False & 0.0058 & 0.0009 & False \\ 
        11 & 81-91 & GRU & -1.6168 & -1.6016 & True & -0.0059 & -0.0056 & True \\ 
        12 & 81-91 & LSTM & -5.002 & -4.9653 & True & -0.0097 & -0.0098 & False \\
        13 & 90-02 & BLSTM & -1.1502 & -1.4366 & False & 0.0036 & 0.0057 & True \\ 
        14 & 90-02 & GRU & -1.2376 & -1.2406 & False & 0.0029 & 0.0061 & True \\ 
        15 & 90-02 & LSTM & -2.2247 & -4.5126 & False & 0.0004 & 0.0113 & True \\ 
        16 & 01-10 & BLSTM & -1.1756 & -0.5414 & True & -0.0015 & 0.0061 & True \\ 
        17 & 01-10 & GRU & -0.3507 & -0.3635 & False & 0.0078 & 0.0028 & False \\ 
        18 & 01-10 & LSTM & -1.1505 & -6.9133 & False & 0.0094 & -0.0361 & False \\
        19 & 07-20 & BLSTM & -0.383 & -1.065 & False & 0.0306 & 0.0443 & True \\ 
        20 & 07-20 & GRU & -0.1678 & -0.8234 & False & 0.0308 & 0.0452 & True \\ 
        21 & 07-20 & LSTM & -3.3818 & -4.4834 & False & -0.0086 & 0.063 & True \\ \hline
        \multicolumn{3}{c}{Count} & 21 & 21 & 6 & 21 & 21 & 14 \\ \hline
\end{xltabular}
\end{singlespace}
\end{small}

Tables \ref{tab:rf_non_rf_trained_mse} and \ref{tab:rf_non_rf_trained_r2_cer} show the results from comparing models trained with and without $r_f$. Comparing models based on their recession MSEs, most models do not show major improvements after the inclusion of $r_f$. A similar conclusion can be drawn from the comparisons of$R^2$ and CER gain for non-recession-trained models. However, a polarising trend was observed for recession-trained (RT) models. We observed a deterioration in $R^2$ values for most RT models after the inclusion of $r_f$ (15 out of 21 models). We also observed an improvement in CER gain for most RT models (14 out of 21 models). This may suggest that $r_f$ reduces relative prediction performance but increases the economic value of an ML model \citep{lu2022cer}. 

\section{Conclusion}\label{sec:d} 

In this work, we have demonstrated that, under optimal hyperparameter fine-tuning, ML models still deliver inferior performance in five out of the seven NBER recession periods when compared to their counterparts' performance in expansion periods. 
We observe that ML model performances are highly correlated with market volatility, represented by the standard deviations of log returns. Indeed, for the two recessions where ML models perform well, their market volatilities are also low. Empirically, we conclude that data shifts occur during most of the past recession periods, exemplified by their increased market volatility. Although the negative impacts of data shifts on ML model performance have been extensively characterized in ML applications \citep{quinonero2008dataset}, there are few studies exploring the data-shifting effects of the business cycle, particularly in the context of stock price forecasting. 
Furthermore, we also explored two measures that aim to reduce the deleterious effects of a recession: 1) the inclusion of recession data in the training set, and 2) the addition of risk-free rate as a predictor. We show that both measures give mixed success in improving model performances. These results suggest that more effort is needed to improve ML model performances with volatile data.

\doublespacing
Moreover, it is worth noting that there is no clear winner among the three classes of RNN-based models we have explored. As summarised in Table \ref{tab:model_type_comparison} below, for most sub-periods the GRU model has the best performance except for 2 sub-periods (80-83 and 07-20) where BLSTM showed higher performance. Across all sub-periods, LSTM had poorer predictions compared to GRU; a finding consistent with other empirical evaluations of RNNs in forecasting \citep{8628641, shahi2020stock, saud2020analysis}. 

\begin{small}
\vspace{-10pt}
\begin{singlespace}
\begin{xltabular}{\textwidth}{L{1}L{1}L{1}L{1}L{1}L{1}L{1}L{1}}  
    \caption[Comparing recession MSEs of Different RNN Models.]{Comparing Recession MSEs of Different RNN Models. For each sub-period, the model with the lowest MSE (best performance) is indicated in bold.} \\ \hline
    \label{tab:model_type_comparison}
    \textbf{Model} & \textbf{69-76} & \textbf{73-80} & \textbf{80-83} & \textbf{81-91} & \textbf{90-02} & \textbf{01-10} & \textbf{07-20} \\   \hline
    \endfirsthead
    \endhead
    LSTM & 0.2469 & 0.0503 & 0.0200 & 0.1207 & 0.2926 & 0.0505 & 1.6498 \\ 
    BLSTM & 0.0600 & 0.0297 & \textbf{0.0081} & 0.0586 & 0.1867 & 0.0461 & \textbf{0.4764} \\ 
    GRU & \textbf{0.0286} & \textbf{0.0268} & 0.0115 & \textbf{0.034} & \textbf{0.1147} & \textbf{0.0345} & 0.6112 \\ \hline
    \end{xltabular}
\end{singlespace}
\vspace{-10pt}
\end{small}

\doublespacing
Furthermore, we can see strong correlations among the three evaluation metrics (Table \ref{tab:metric_correlations}) with MSE having the highest correlation with log-return volatility. $R^2$ was the least correlated as it normalises model performance according to prediction difficulty. The highly positive relationship between CERG and log-return standard deviation further suggests that ML models become more valuable during volatile periods.

\begin{small}
\begin{singlespace}
\begin{xltabular}{\textwidth}{L{1.9}L{0.7}L{0.7}L{0.7}}
    \caption[Pearson's correlation coefficient of each metric with log-return standard deviation.]{Pearson's correlation coefficient of each metric with log-return standard deviation. } \\ \hline
    \label{tab:metric_correlations}
    ~ & \textbf{MSE} & \textbf{$R^2$} & \textbf{CER Gain} \\ \hline
    \endfirsthead
    \endhead
    \textbf{Correlation With Log-Return S.D.} & 0.744147 & 0.214457 & 0.641578 \\ \hline
    \end{xltabular}
\end{singlespace}
\end{small}

\begin{figure}[htbp]
\centering
\includegraphics[trim=5 7 10 45,clip,width=\textwidth]{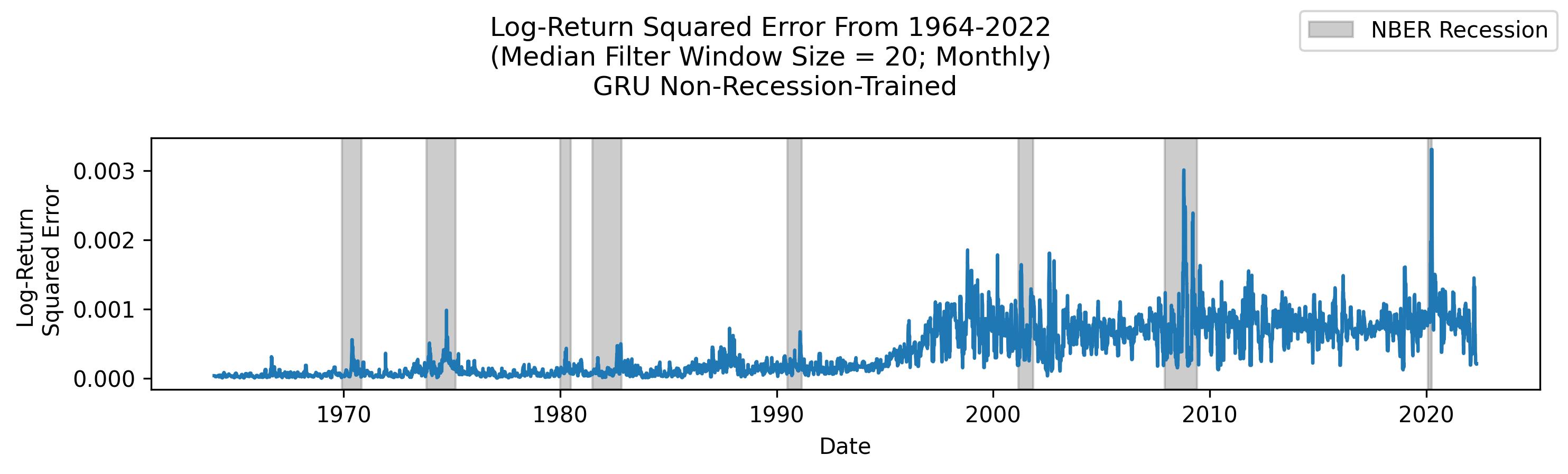}
\caption[Graph showing the log-return squared errors of GRU Non-Recession-Trained models on data from 1964-2022]{Squared errors of GRU Non-Recession-Trained models on data from 1964-2022. Predicted and true stock prices were converted to predicted and true log returns to detrend the stock prices before squared error calculation. Square errors were passed through a median filter with a window size of 20 (number of days in a month) to reduce noise. During a recession (Grey-shaded regions), peaks in squared error can be observed. }
\label{fig:squared_error_plot}
\vspace{-10pt}
\end{figure}

Another interesting observation from our findings is that we have demonstrated the volatility during recessions universally affects stock price predictions negatively, regardless of the feature set used or the presence of recession in the in-sample set. A useful extension of this finding as a data-driven recession index is depicted in Figure~\ref{fig:squared_error_plot}. By analysing the squared errors of an ML model, we can identify recessions without referring to macroeconomic measures, which may be delayed. 

From the results of this study, we recommend ML practitioners in finance evaluate their ML models during both recession and non-recession periods to better understand the effectiveness of their model throughout the business cycle. 





\vspace{-10pt}
\begin{singlespace}
\bibliographystyle{elsarticle-harv}%
\bibliography{citations}

\begin{thebibliography}{28}
\expandafter\ifx\csname natexlab\endcsname\relax\def\natexlab#1{#1}\fi
\providecommand{\url}[1]{\texttt{#1}}
\providecommand{\href}[2]{#2}
\providecommand{\path}[1]{#1}
\providecommand{\DOIprefix}{doi:}
\providecommand{\ArXivprefix}{arXiv:}
\providecommand{\URLprefix}{URL: }
\providecommand{\Pubmedprefix}{pmid:}
\providecommand{\doi}[1]{\href{http://dx.doi.org/#1}{\path{#1}}}
\providecommand{\Pubmed}[1]{\href{pmid:#1}{\path{#1}}}
\providecommand{\bibinfo}[2]{#2}
\ifx\xfnm\relax \def\xfnm[#1]{\unskip,\space#1}\fi
\bibitem[{Agarap(2018)}]{agarap2018deep}
\bibinfo{author}{Agarap, A.F.}, \bibinfo{year}{2018}.
\newblock \bibinfo{title}{Deep learning using rectified linear units (relu)}.
\newblock \bibinfo{journal}{arXiv preprint arXiv:1803.08375} .
\bibitem[{Bianchi et~al.(2021)Bianchi, B{\"u}chner and
  Tamoni}]{bianchi2021bond}
\bibinfo{author}{Bianchi, D.}, \bibinfo{author}{B{\"u}chner, M.},
  \bibinfo{author}{Tamoni, A.}, \bibinfo{year}{2021}.
\newblock \bibinfo{title}{Bond risk premiums with machine learning}.
\newblock \bibinfo{journal}{Review of Financial Studies} \bibinfo{volume}{34},
  \bibinfo{pages}{1046--1089}.
\bibitem[{Campbell and Thompson(2008)}]{campbell2008oosrsquared}
\bibinfo{author}{Campbell, J.Y.}, \bibinfo{author}{Thompson, S.B.},
  \bibinfo{year}{2008}.
\newblock \bibinfo{title}{Predicting excess stock returns out of sample: Can
  anything beat the historical average?}
\newblock \bibinfo{journal}{Review of Financial Studies} \bibinfo{volume}{21},
  \bibinfo{pages}{1509--1531}.
\bibitem[{Cao et~al.(2020)Cao, Zhu, Wang, Demazeau and Zhang}]{cao2020deep}
\bibinfo{author}{Cao, W.}, \bibinfo{author}{Zhu, W.}, \bibinfo{author}{Wang,
  W.}, \bibinfo{author}{Demazeau, Y.}, \bibinfo{author}{Zhang, C.},
  \bibinfo{year}{2020}.
\newblock \bibinfo{title}{A deep coupled lstm approach for usd/cny exchange
  rate forecasting}.
\newblock \bibinfo{journal}{IEEE Intelligent Systems} \bibinfo{volume}{35},
  \bibinfo{pages}{43--53}.
\bibitem[{Chauvet et~al.(2013)Chauvet, G{\"o}tz and Hecq}]{chauvet2013realized}
\bibinfo{author}{Chauvet, M.}, \bibinfo{author}{G{\"o}tz, T.},
  \bibinfo{author}{Hecq, A.}, \bibinfo{year}{2013}.
\newblock \bibinfo{title}{Realized volatility and business cycle fluctuations:
  A mixed-frequency VAR approach}.
\newblock \bibinfo{type}{Technical Report}. Working paper, University of
  California Riverside and Maastricht University.
\bibitem[{Cho et~al.(2014)Cho, Van~Merri{\"e}nboer, Bahdanau and
  Bengio}]{cho2014properties}
\bibinfo{author}{Cho, K.}, \bibinfo{author}{Van~Merri{\"e}nboer, B.},
  \bibinfo{author}{Bahdanau, D.}, \bibinfo{author}{Bengio, Y.},
  \bibinfo{year}{2014}.
\newblock \bibinfo{title}{On the properties of neural machine translation:
  Encoder-decoder approaches}.
\newblock \bibinfo{journal}{arXiv preprint arXiv:1409.1259} .
\bibitem[{Choudhry et~al.(2016)Choudhry, Papadimitriou and
  Shabi}]{choudhry2016stock}
\bibinfo{author}{Choudhry, T.}, \bibinfo{author}{Papadimitriou, F.I.},
  \bibinfo{author}{Shabi, S.}, \bibinfo{year}{2016}.
\newblock \bibinfo{title}{Stock market volatility and business cycle: Evidence
  from linear and nonlinear causality tests}.
\newblock \bibinfo{journal}{Journal of Banking \& Finance}
  \bibinfo{volume}{66}, \bibinfo{pages}{89--101}.
\bibitem[{Corradi et~al.(2013)Corradi, Distaso and
  Mele}]{corradi2013macroeconomic}
\bibinfo{author}{Corradi, V.}, \bibinfo{author}{Distaso, W.},
  \bibinfo{author}{Mele, A.}, \bibinfo{year}{2013}.
\newblock \bibinfo{title}{Macroeconomic determinants of stock volatility and
  volatility premiums}.
\newblock \bibinfo{journal}{Journal of Monetary Economics}
  \bibinfo{volume}{60}, \bibinfo{pages}{203--220}.
\bibitem[{Diebold and Yilmaz(2008)}]{diebold2008macroeconomic}
\bibinfo{author}{Diebold, F.X.}, \bibinfo{author}{Yilmaz, K.},
  \bibinfo{year}{2008}.
\newblock \bibinfo{title}{Macroeconomic volatility and stock market volatility,
  worldwide}.
\newblock \bibinfo{type}{Technical Report}. National Bureau of Economic
  Research.
\bibitem[{Engle and Rangel(2008)}]{engle2008spline}
\bibinfo{author}{Engle, R.F.}, \bibinfo{author}{Rangel, J.G.},
  \bibinfo{year}{2008}.
\newblock \bibinfo{title}{The spline-garch model for low-frequency volatility
  and its global macroeconomic causes}.
\newblock \bibinfo{journal}{The review of financial studies}
  \bibinfo{volume}{21}, \bibinfo{pages}{1187--1222}.
\bibitem[{Fama(1990)}]{fama1990stock}
\bibinfo{author}{Fama, E.F.}, \bibinfo{year}{1990}.
\newblock \bibinfo{title}{Stock returns, expected returns, and real activity}.
\newblock \bibinfo{journal}{The journal of finance} \bibinfo{volume}{45},
  \bibinfo{pages}{1089--1108}.
\bibitem[{Fama and French(1989)}]{fama1989business}
\bibinfo{author}{Fama, E.F.}, \bibinfo{author}{French, K.R.},
  \bibinfo{year}{1989}.
\newblock \bibinfo{title}{Business conditions and expected returns on stocks
  and bonds}.
\newblock \bibinfo{journal}{Journal of financial economics}
  \bibinfo{volume}{25}, \bibinfo{pages}{23--49}.
\bibitem[{Fama and French(1993)}]{fama1993common}
\bibinfo{author}{Fama, E.F.}, \bibinfo{author}{French, K.R.},
  \bibinfo{year}{1993}.
\newblock \bibinfo{title}{Common risk factors in the returns on stocks and
  bonds}.
\newblock \bibinfo{journal}{Journal of Financial Economics}
  \bibinfo{volume}{33}, \bibinfo{pages}{3--56}.
\bibitem[{Fama and French(2017)}]{fama2017choosing}
\bibinfo{author}{Fama, E.F.}, \bibinfo{author}{French, K.R.},
  \bibinfo{year}{2017}.
\newblock \bibinfo{title}{Choosing factors}.
\newblock \bibinfo{journal}{Fama-Miller Working Paper, Tuck School of Business
  Working Paper} , \bibinfo{pages}{16--17}.
\bibitem[{Graves and Schmidhuber(2005)}]{1556215}
\bibinfo{author}{Graves, A.}, \bibinfo{author}{Schmidhuber, J.},
  \bibinfo{year}{2005}.
\newblock \bibinfo{title}{Framewise phoneme classification with bidirectional
  lstm networks}, in: \bibinfo{booktitle}{Proceedings. 2005 IEEE International
  Joint Conference on Neural Networks, 2005.}, pp. \bibinfo{pages}{2047--2052
  vol. 4}.
\newblock \DOIprefix\doi{10.1109/IJCNN.2005.1556215}.
\bibitem[{Gu et~al.(2020)Gu, Kelly and Xiu}]{guetal2020}
\bibinfo{author}{Gu, S.}, \bibinfo{author}{Kelly, B.}, \bibinfo{author}{Xiu,
  D.}, \bibinfo{year}{2020}.
\newblock \bibinfo{title}{Empirical asset pricing via machine learning}.
\newblock \bibinfo{journal}{Review of Financial Studies} \bibinfo{volume}{33},
  \bibinfo{pages}{2223--2273}.
\bibitem[{Gu et~al.(2021)Gu, Kelly and Xiu}]{gu2021autoencoder}
\bibinfo{author}{Gu, S.}, \bibinfo{author}{Kelly, B.}, \bibinfo{author}{Xiu,
  D.}, \bibinfo{year}{2021}.
\newblock \bibinfo{title}{Autoencoder asset pricing models}.
\newblock \bibinfo{journal}{Journal of Econometrics} \bibinfo{volume}{222},
  \bibinfo{pages}{429--450}.
\bibitem[{Hamayel and Owda(2021)}]{hamayel2021novel}
\bibinfo{author}{Hamayel, M.J.}, \bibinfo{author}{Owda, A.Y.},
  \bibinfo{year}{2021}.
\newblock \bibinfo{title}{A novel cryptocurrency price prediction model using
  gru, lstm and bi-lstm machine learning algorithms}.
\newblock \bibinfo{journal}{AI} \bibinfo{volume}{2}, \bibinfo{pages}{477--496}.
\bibitem[{Hamilton and Lin(1996)}]{hamilton1996stock}
\bibinfo{author}{Hamilton, J.D.}, \bibinfo{author}{Lin, G.},
  \bibinfo{year}{1996}.
\newblock \bibinfo{title}{Stock market volatility and the business cycle}.
\newblock \bibinfo{journal}{Journal of applied econometrics}
  \bibinfo{volume}{11}, \bibinfo{pages}{573--593}.
\bibitem[{Heaton et~al.(2017)Heaton, Polson and
  Witte}]{heaton2017portfolioselection}
\bibinfo{author}{Heaton, J.B.}, \bibinfo{author}{Polson, N.G.},
  \bibinfo{author}{Witte, J.H.}, \bibinfo{year}{2017}.
\newblock \bibinfo{title}{Deep learning for finance: deep portfolios}.
\newblock \bibinfo{journal}{Applied Stochastic Models in Business and Industry}
  \bibinfo{volume}{33}, \bibinfo{pages}{3--12}.
\bibitem[{Hochreiter and Schmidhuber(1997)}]{hochreiter1997long}
\bibinfo{author}{Hochreiter, S.}, \bibinfo{author}{Schmidhuber, J.},
  \bibinfo{year}{1997}.
\newblock \bibinfo{title}{Long short-term memory}.
\newblock \bibinfo{journal}{Neural computation} \bibinfo{volume}{9},
  \bibinfo{pages}{1735--1780}.
\bibitem[{Hossain et~al.(2018)Hossain, Karim, Thulasiram, Bruce and
  Wang}]{8628641}
\bibinfo{author}{Hossain, M.A.}, \bibinfo{author}{Karim, R.},
  \bibinfo{author}{Thulasiram, R.}, \bibinfo{author}{Bruce, N.D.B.},
  \bibinfo{author}{Wang, Y.}, \bibinfo{year}{2018}.
\newblock \bibinfo{title}{Hybrid deep learning model for stock price
  prediction}, in: \bibinfo{booktitle}{2018 IEEE Symposium Series on
  Computational Intelligence (SSCI)}, pp. \bibinfo{pages}{1837--1844}.
\newblock \DOIprefix\doi{10.1109/SSCI.2018.8628641}.
\bibitem[{Lu and Han(2022)}]{lu2022cer}
\bibinfo{author}{Lu, Y.J.}, \bibinfo{author}{Han, Y.}, \bibinfo{year}{2022}.
\newblock \bibinfo{title}{Macroeconomic extrapolation, machine learning, and
  equity risk premium forecast}.
\newblock \bibinfo{journal}{Available at SSRN 4102419} .
\bibitem[{Quinonero-Candela et~al.(2008)Quinonero-Candela, Sugiyama,
  Schwaighofer and Lawrence}]{quinonero2008dataset}
\bibinfo{author}{Quinonero-Candela, J.}, \bibinfo{author}{Sugiyama, M.},
  \bibinfo{author}{Schwaighofer, A.}, \bibinfo{author}{Lawrence, N.D.},
  \bibinfo{year}{2008}.
\newblock \bibinfo{title}{Dataset shift in machine learning}.
\newblock \bibinfo{publisher}{Mit Press}.
\bibitem[{Saud and Shakya(2020)}]{saud2020analysis}
\bibinfo{author}{Saud, A.S.}, \bibinfo{author}{Shakya, S.},
  \bibinfo{year}{2020}.
\newblock \bibinfo{title}{Analysis of look back period for stock price
  prediction with rnn variants: A case study on banking sector of nepse}.
\newblock \bibinfo{journal}{Procedia Computer Science} \bibinfo{volume}{167},
  \bibinfo{pages}{788--798}.
\bibitem[{Schwert(1989)}]{schwert1989business}
\bibinfo{author}{Schwert, G.W.}, \bibinfo{year}{1989}.
\newblock \bibinfo{title}{Business cycles, financial crises, and stock
  volatility}, in: \bibinfo{booktitle}{Carnegie-Rochester Conference series on
  public policy}, \bibinfo{organization}{Elsevier}. pp.
  \bibinfo{pages}{83--125}.
\bibitem[{Seabe et~al.(2023)Seabe, Moutsinga and Pindza}]{seabe2023forecasting}
\bibinfo{author}{Seabe, P.L.}, \bibinfo{author}{Moutsinga, C.R.B.},
  \bibinfo{author}{Pindza, E.}, \bibinfo{year}{2023}.
\newblock \bibinfo{title}{Forecasting cryptocurrency prices using lstm, gru,
  and bi-directional lstm: A deep learning approach}.
\newblock \bibinfo{journal}{Fractal and Fractional} \bibinfo{volume}{7},
  \bibinfo{pages}{203}.
\bibitem[{Shahi et~al.(2020)Shahi, Shrestha, Neupane and Guo}]{shahi2020stock}
\bibinfo{author}{Shahi, T.B.}, \bibinfo{author}{Shrestha, A.},
  \bibinfo{author}{Neupane, A.}, \bibinfo{author}{Guo, W.},
  \bibinfo{year}{2020}.
\newblock \bibinfo{title}{Stock price forecasting with deep learning: A
  comparative study}.
\newblock \bibinfo{journal}{Mathematics} \bibinfo{volume}{8},
  \bibinfo{pages}{1441}.

\end{thebibliography}
\end{singlespace}

\end{document}